\documentclass[aps,pra,twocolumn,groupedaddress,floatfix,showpacs,nofootinbib]{revtex4}
\usepackage{graphicx}
\usepackage{amsmath}
\usepackage{bm}
\usepackage{amstext}
\usepackage{amsxtra}
\usepackage{placeins}
\usepackage{float}

\usepackage{times}

\begin{document}

\newcommand{\To}{T_c^0}
\newcommand{\kB}{k_{\rm B}}
\newcommand{\dT}{\Delta T_c}
\newcommand{\lo}{\lambda_0}
\newcommand{\cs}{$\clubsuit \; $}
\newcommand{\thold}{t_{\rm hold}}
\newcommand{\Nmf}{N_c^{\rm MF}}
\newcommand{\Tmf}{T_c^{\rm MF}}
\newcommand{\bra}[1]{\langle #1|}
\newcommand{\ket}[1]{|#1\rangle}
\newcommand{\downstate}{\left\vert\downarrow\right\rangle}
\newcommand{\upstate}{\left\vert\uparrow\right\rangle}
\newcommand{\vF}{v_{c}^{\rm F}}
\newcommand{\vs}{v_{c}^{\rm s}}

\title{
Quantised supercurrent decay in an annular Bose-Einstein condensate
}

\author{Stuart Moulder, Scott Beattie, Robert P. Smith, Naaman Tammuz, and Zoran Hadzibabic}
\affiliation{Cavendish Laboratory, University of Cambridge, J.~J.~Thomson Ave., Cambridge CB3~0HE, United Kingdom}

\begin{abstract}
We study the metastability and decay of multiply-charged superflow in a ring-shaped atomic Bose-Einstein condensate. Supercurrent corresponding to a giant vortex with topological charge up to $q=10$ is phase-imprinted optically and detected  both interferometrically and kinematically.
We observe $q=3$ superflow persisting for up to a minute and clearly resolve a cascade of quantised steps in its decay.  These stochastic decay events, associated with vortex-induced $2 \pi$ phase slips, correspond to collective jumps of atoms between discrete $q$ values. 
We demonstrate the ability to detect quantised rotational states with $> 99\,\%$ fidelity, which allows a detailed quantitative study of time-resolved phase-slip dynamics. We find that the supercurrent decays rapidly if the superflow speed exceeds a critical velocity in good agreement with numerical simulations, and we also observe rare stochastic phase slips for superflow speeds below the critical velocity.

\end{abstract}

\date{\today}


%03.75.Hh	Static properties of condensates; thermodynamical, statistical, and structural properties
%67.85.-d 	Ultracold gases, trapped gases
%03.75.Kk 	Dynamic properties of condensates; collective and hydrodynamic excitations, superfluid flow
%Optical angular momentum (quantum optics), 42.50.Tx
%47.37.+q 	Hydrodynamic aspects of superfluidity; quantum fluids
%67.85.De 	Dynamic properties of condensates; excitations, and superfluid flow
%37.10.Vz 	Mechanical effects of light on atoms, molecules, and ions

\maketitle

\section{Introduction}

\begin{figure}[bp]
\includegraphics[width=0.9\columnwidth]{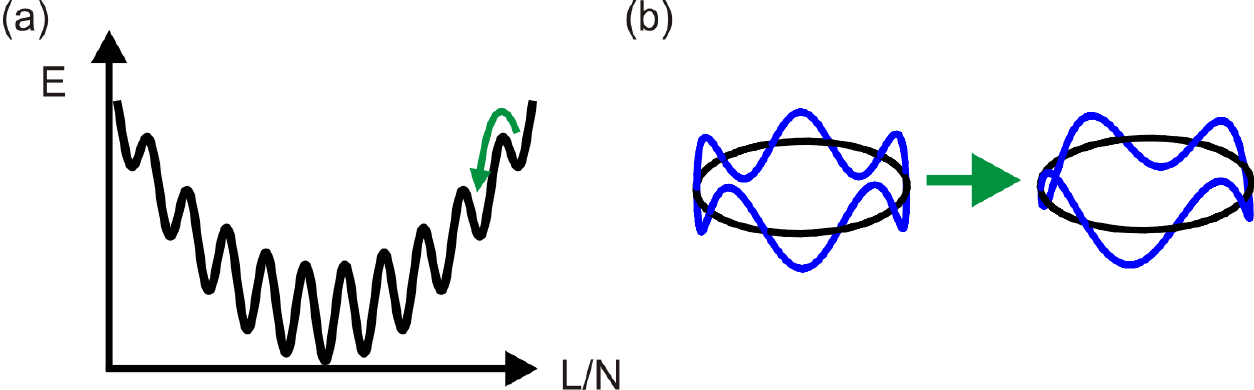}
\caption{(Color online) Metastability and decay of supercurrents. (a) Energy landscape of a ring-shaped superfluid. Local minima correspond to metastable states with quantised angular momentum per particle, $L/N = q\hbar$. (b) Decay between the discrete
 $q$ states involves a vortex-mediated phase slip, illustrated here for $q = 5 \rightarrow 4$.}
\label{fig:1}
\end{figure}

Superfluid flow of a Bose-Einstein condensate (BEC) in a multiply-connected ring geometry is the archetypal metastable many-body state.
The phase of the macroscopic BEC wave function must wind around the ring by an integer multiple of $2 \pi$, corresponding to the charge $q$ of a vortex trapped inside the ring.  Macroscopic states with different $q$ values are topologically distinct and separated by energy barriers [see Fig.~\ref{fig:1}(a)].
Consequently, although the true ground state of the system in a non-rotating trap is $q=0$, a $q \neq 0$ supercurrent can be extremely long-lived, and largely immune to perturbations such as disorder and thermal fluctuations. Stability and decay of supercurrents have been studied for decades in helium superfluids \cite{Feynman:1955,Reppy:1965,Avenel:1985,Varoquaux:1986,Davis:1992} and thin-wire superconductors \cite{Little:1967,Langer:1967,McCumber:1970,Sahu:2009, Li:2011}, but the decay process is still not fully understood \cite{Halperin:2010}.
A ring-shaped superfluid was also proposed as the ideal laboratory system for simulation of pulsar glitches \cite{Anderson:1975}, associated with jumps in the rotation of the superfluid neutron star interior \cite{Packard:1972, Anderson:1975}.

Atomic BECs trapped in a ring geometry \cite{Sauer:2001,Gupta:2005,Arnold:2006,Ryu:2007,Henderson:2009,Ramanathan:2011,Sherlock:2011} are attractive both for fundamental studies of superfluidity and for applications in interferometry \cite{Gustavson:1997,Halkyard:2010} and atomtronics \cite{Seaman:2007}. Recently, $q=1$ superflow persisting for 40 s was observed  and 
studies of flow through a weak link created by a potential barrier revealed a relatively sharply defined superflow critical velocity, $v_c$ \cite{Ramanathan:2011}. The observed $v_c$ was consistent with the Feynman estimate, $\vF$, strongly suggesting a vortex-induced phase slip [see Fig.~\ref{fig:1}(b)] as the dominant supercurrent decay mechanism.

While a $q=1$ vortex can persist for several seconds even in a simply-connected BEC, any $q>1$ vortex is fundamentally unstable in such a geometry \cite{Shin:2004b,Ryu:2007}. In a ring trap, a $q=2$ vortex was observed to survive for at least 0.5 s \cite{Ryu:2007}. However, the decay of $q>1$ vortices in a multiply-connected geometry has not yet been studied.

In this paper, we demonstrate and study extreme metastability of multiply-charged superflow in an annular BEC.
Using optical phase-imprinting \cite{Andersen:2006} we prepare annular BECs in metastable rotational states corresponding to vortex charges up to $q=10$. To quantitatively study the supercurrent decay with sufficient statistics we focus on condensates initially prepared in a $q=3$ state.
We observe $q=3$ superflow persisting for up to a minute in a multiply-connected trap, and explicitly show that the supercurrent is quantised. The cascade of quantised decay steps unambiguously confirms $2 \pi$ phase slips as the supercurrent decay mechanism. We demonstrate the ability to read-out quantised rotational states with $> 99\, \%$ fidelity, which opens the possibility to quantitatively study the dynamics of phase-slips for different superflow speeds.
While a rapid $q \rightarrow q-1$ decay occurs if the flow speed $v_s (q)$ reaches a critical velocity, we also observe more rare stochastic phase slips for $v_s < v_c$.
After each phase slip the system re-stabilises in a lower metastable rotational state.
We find that the critical velocities for different $q$ states are of the same order of magnitude as the Feynman estimate, but are more quantitatively predicted by a numerical simulation linking $v_c$ to the condensate surface instability against a vortex penetrating the annulus \cite{Anglin:2001, Dubessy:2012}.

The paper is divided into six sections. In Sec.~\ref{sec:theory} we outline the theoretical considerations concerning the stability and decay of supercurrents in annular atomic superfluids. In Sec.~\ref{sec:exp} we describe our preparation and detection  of metastable supercurrents.  In Sec.~\ref{sec:quantisation} we present our observations of quantised supercurrent decay and long-lived multiply-charged superflow. 
In Sec.~\ref{sec:dynamics} we discuss the phase-slip dynamics for different superflow speeds.
Finally, we summarise our results and briefly discuss future research directions in Sec.~\ref{sec:conclusion}.

\section{Supercurrents in annular condensates}
\label{sec:theory}

\subsection{Topologically protected superflow states}

The physical origin of the supercurrent metastability is qualitatively illustrated in Fig.~\ref{fig:1}(a).
For $N$ atoms held in a ring trap, the average angular momentum per particle in general need not be quantised, but for a superfluid gas such quantisation is energetically preferred. The ``parabolic washboard" landscape depicts the energy $E$ of a superfluid system for different fixed values of the total angular momentum $L$ \cite{F}. The local minima of $E$
correspond to topologically distinct metastable states with $L/N =q\hbar$. A direct $\Delta q=1$ transition between two such minima involves a discontinuous $2 \pi$ phase slip in the condensate wave function, occurring when a singly-charged vortex crosses the annulus.

More generally, a superfluid can in principle also shed angular momentum in ways that break the $L/N$ quantisation, including condensate fragmentation and collective excitations such as solitons \cite{Kanamoto:2008}.
For low enough superflow speeds all such processes are energetically costly and suppressed to various degrees.
The dominant superflow decay mechanism depends on the system's dimensions, temperature, and the strength of interactions  \cite{McCumber:1970, Kanamoto:2008, Halperin:2010}, and is often difficult to predict.

The dissipative supercurrent decay is strictly speaking always stochastic, even if (for sufficiently high flow speeds) the superflow is unstable in the thermodynamic sense. However, we can distinguish qualitatively different decay regimes:

(1) If $v_s$ exceeds the critical velocity for some decay process, the decay becomes likely. Ultimately it can occur on some microscopic timescale, which for an atomic BEC is in the millisecond range. In this case, from an experimental point of view the decay can appear essentially instantaneous and deterministic. For example, we can not talk about a persistent current if it ``persists" for much less than one rotation period ($\sim 300\;$ms in our experiments). 
%We can thus empirically define such decay as ``rapid." 
%In our experiments this corresponds to a timescale of $\sim 300\;$ms.

(2) For $v_s \ll v_c$ the decay is strongly suppressed and the superflow can be almost perfectly stable, as for example observed in bulk superconductors. 

(3) In between these two extremes, metastable superflow should persist for much longer than the characteristic microscopic timescale of the physical system, but rare stochastic decay events can still occur through quantum or thermal fluctuations \cite{Langer:1967,McCumber:1970,McKay:2008,Sahu:2009,Li:2011}.
Such stochastic phase slips are, for example, associated with the residual resistance in thin-wire superconductors 
\cite{Halperin:2010}. 

\subsection{Critical velocity for vortex-induced phase slips}

The critical velocity for the occurrence of vortex-induced phase slips was famously first estimated by Feynman: 
\begin{equation}
\vF = \frac{\hbar}{mr} \ln \left( \frac{r}{\xi} \right) \; ,
\label{eq:Feynman}
\end{equation} 
where $m$ is the atom mass, $r$ the annulus width, and $\xi$ the healing length. 
This estimate is based on general energetic arguments, namely the cost of a vortex crossing a high superfluid-density region of characteristic size $r$. It does not take into account the dynamical effects associated with the vortex penetrating the BEC, nor the details of geometry such as the variation of the condensate density due to the harmonic trapping along the directions transverse to the ring.
It is also important to note that in Feynman's theory $\vF$ does not correspond to a sharp boundary between stable and unstable superflow. Rather, $\vF$ just sets the natural scale for the superflow speed $v_s$ at which phase-slip induced supercurrent decay should become energetically favourable.
Neverthless, in some cases Eq.~(\ref{eq:Feynman}) is found to provide a good estimate of $v_c$ \cite{Ramanathan:2011}.

In experiments on simply-connected rotating atomic gases \cite{Madison:2000a, AboShaeer:2001, Madison:2001, Raman:2001b, Zwierlein:2005} it was often found that the critical velocity for a vortex entering the condensate was higher than predicted purely by global energetic arguments. This higher $v_c$ is associated with dynamical instabilities of surface excitations, which provide the necessary microscopic route for vortex nucleation.
The ``surface" critical velocity at which such instability occurs was derived by Anglin \cite{Anglin:2001}, properly taking into account the variation of the condensate density near its edge:
\begin{equation}
\vs =  \sqrt{\frac{2 \hbar \omega}{m}} \left( \frac{\mu}{\hbar \omega} \right)^{1/6} \; ,
\label{eq:surface}
\end{equation}
where $\omega$ is the radial trapping frequency (along the direction perpendicular to the rotation axis) and $\mu$ the chemical potential. The arguments of Ref.~\cite{Anglin:2001} are local, and consider only a surface region of size several $\xi$. Hence the theory should be equally applicable to rotating annular condensates, as long as both the width of the annulus and its inner radius are much larger than $\xi$. Indeed the theory of~\cite{Anglin:2001} was recently extended to ring geometry by Dubessy {\it et al.}~\cite{Dubessy:2012}. In this case the expression for $\vs$ is the same, but one notes that due to the nature of the superfluid flow with quantised angular momentum the critical velocity is always first reached at the inner surface of the annulus. In other words, while a phase slip can formally be thought of either as a vortex crossing the annulus to exit the ring, or an anti-vortex entering the ring, in reality the former process is always more likely.

In our experiments $\vs$ is always higher than $\vF$ and the geometric criteria for the applicability of Eq.~(\ref{eq:surface})  are satisfied.  We will address the comparison of our observations with the two theories of phase-slip $v_c$ in Sec.~\ref{sec:dynamics}, after introducing our experimental methods (Sec.~\ref{sec:exp}) and showing that in our experiments vortex-induced phase slips are indeed the relevant supercurrent decay mechanism (Sec.~\ref{sec:quantisation}).

\section{Preparation and detection of supercurrent}
\label{sec:exp}

\begin{figure}[tbp]
\includegraphics[width=0.9\columnwidth]{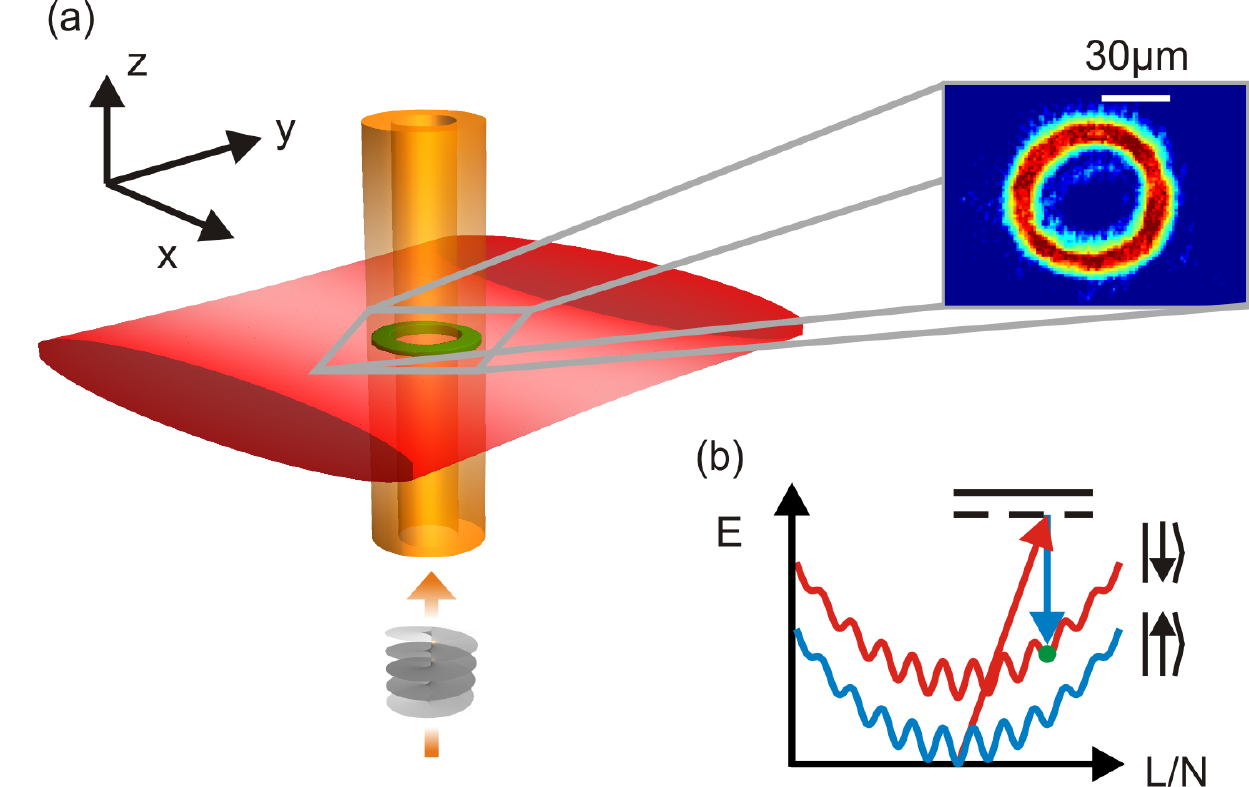}
\caption{(Color online) Preparation of metastable supercurrent in an annular condensate.  (a) The optical ring trap is created by intersecting a horizontal ``sheet" laser beam with a vertical ``tube" LG$^{\ell}$ beam; the absorption image shows a BEC in an $\ell=10$ trap. (b) Two-photon Raman transfer of atoms into a metastable $q=\ell$ state is achieved using the LG$^{\ell}$ trapping beam (red) and a co-propagating Gaussian beam (blue). An atom undergoing an internal state transfer, $\upstate \rightarrow \downstate$, also absorbs angular momentum $\ell\hbar$ from the  LG$^{\ell}$ laser beam.}
\label{fig:2}
\end{figure}

In our experiments we use a hollow Laguerre-Gauss (LG) mode of an infrared (805 nm) laser beam to both trap the superfluid in a ring geometry and set it into rotation (see Fig.~\ref{fig:2}). In an LG$^\ell$ laser mode each photon carries orbital angular momentum $\ell \hbar$, which can be transferred to an atom via a two-photon Raman process \cite{Andersen:2006}. To prepare different $q$ rotational states we create LG beams with $\ell$ values up to $\ell=10$, using a phase-imprinting spatial light modulator (SLM) \cite{Curtis:2003}.

We start by producing a quasi-pure BEC of $^{87}$Rb atoms \cite{Campbell:2010} and loading it into the ring trap formed at the intersection of a vertical LG beam and a horizontal ``sheet" beam of wavelength 1070 nm [see Fig.~\ref{fig:2}(a)].
To avoid inducing rotation of the BEC during loading into the ring trap, this transfer is done very slowly over 5 s.
We load $\approx 2 \times 10^5$ condensed atoms into the ring, and at no time during the experiment observe a discernible thermal fraction of the gas \cite{systematics}.
The sheet beam
provides a nearly isotropic trapping potential in the $xy$ plane, with trapping frequencies of $6$, $7$, and $400$ Hz along the $\hat{x}$, $\hat{y}$, and $\hat{z}$ directions, respectively. The depth of the ring trap, $V_r$, is set by the power of the LG beam. For $\ell =3$, the ring radius is $\approx 12 \;\mu$m and the radial trapping frequency varies between 75 Hz and 190 Hz for the $V_r$ values used in our experiments. For higher $\ell$ the trap radius increases approximately linearly \cite{LG}.

To set the superfluid into rotation via a two-photon Raman transition, we briefly ($\sim 200$ $\mu$s) pulse on an auxiliary 805 nm Gaussian beam, co-propagating with the trapping LG beam.  As illustrated in Fig. \ref{fig:2}(b), the atoms are transferred between two internal atomic states, $\upstate$ and $\downstate$, and simultaneously pick up angular momentum $\ell \hbar$. The $\upstate$ and $\downstate$ are two Zeeman levels of the $F= 1$ hyperfine ground state, $m_F = 1$ and 0, respectively. The $m_F = -1$ state is detuned from the Raman resonance by the quadratic Zeeman shift in an external magnetic field of 10 gauss.

\begin{figure}[btp]
\includegraphics[width=0.9\columnwidth]{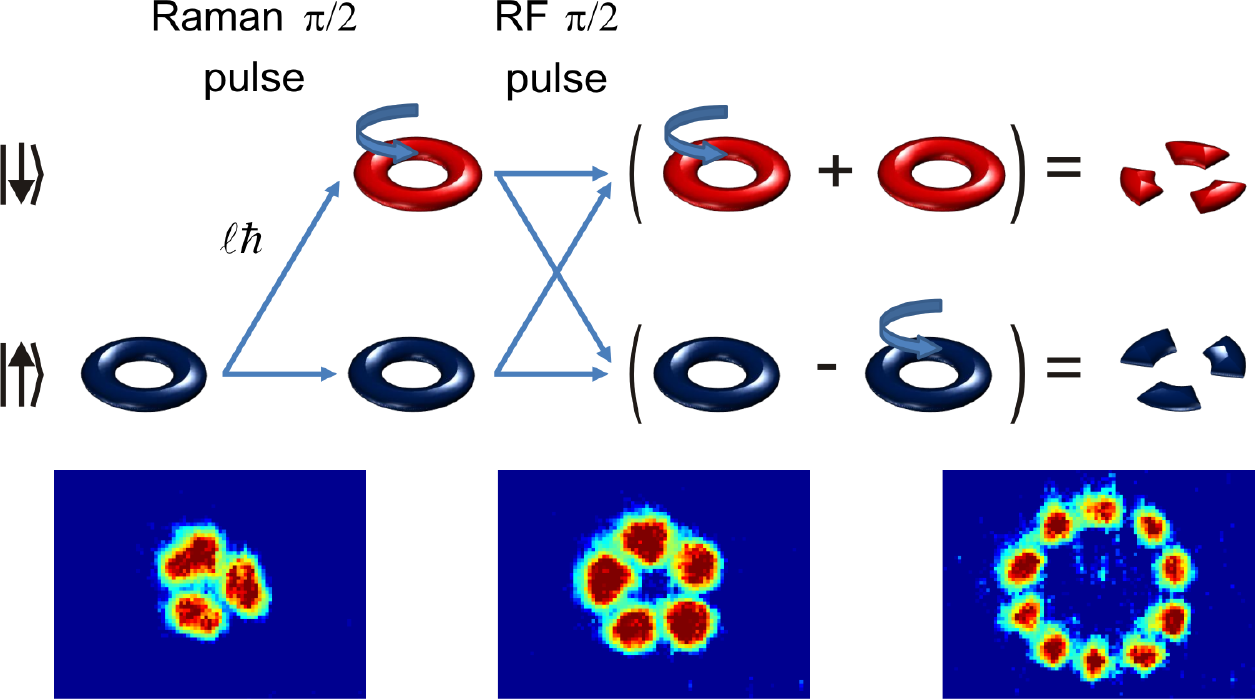}
\caption{(Color online) Interferometric detection of the imprinted phase winding. A combination of Raman and RF $\pi/2$ pulses results in matter-wave interference between stationary and moving atoms, with the number of density peaks equal to $\ell$. Absorption images of the $\upstate$ state, taken 3 ms after releasing the atoms from the trap, show matter-wave interference for $\ell=3, 5,$ and $10$.}
\label{fig:3}
\end{figure}

We first perform a set of interferometric experiments in order to verify the optically imprinted phase winding (see also \cite{Matthews:1999, Inouye:2001,Chevy:2001}). As depicted in Fig. \ref{fig:3}, we apply a $\pi/2$ Raman pulse  which coherently transfers only half the population into the rotating $\downstate$ state.  A subsequent $\pi/2$ radio-frequency (RF) pulse, which carries no angular momentum, mixes the $\upstate$ and $\downstate$ states so that in each spin state we get an interference of rotating ($q=\ell$) and non-rotating ($q=0$) atoms. This matter-wave interference converts the phase winding into a density modulation, with the number of density peaks around the ring equal to $\ell$.  In Fig.~\ref{fig:3} we show the observed interference patterns for $\ell = 3$, 5 and 10.

\begin{figure} [bp]
\includegraphics[width=0.9\columnwidth]{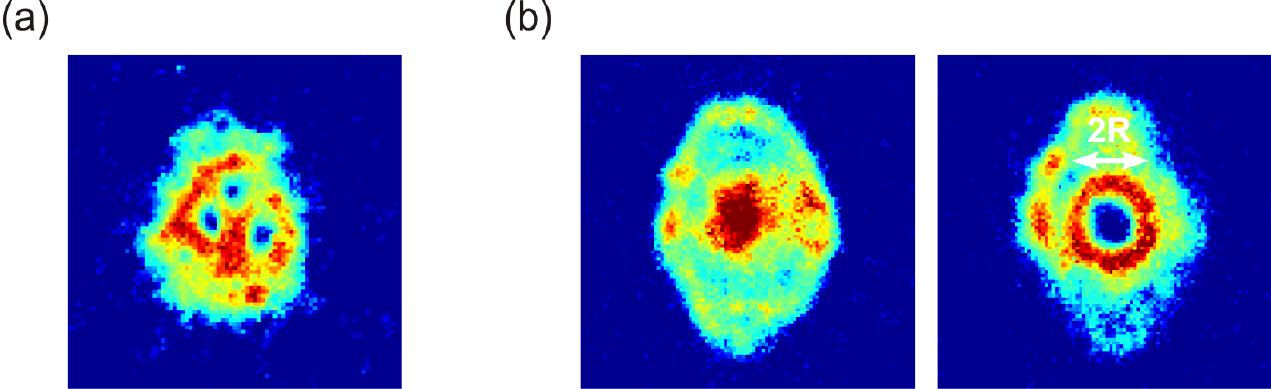}
\caption{(Color online) Detection of superflow. (a) If the ring trap is transformed into a simply-connected sheet trap, the $q=3$ vortex breaks up into $3$ individual vortices. 
(b) Absorption images of non-rotating (left) and rotating (right) BECs after 29 ms of TOF expansion from the ring trap.
We use the radius $R$ to quantify the rotation of the cloud.}
\label{fig:4}
\end{figure}

For our main studies (Sections~\ref{sec:quantisation} and \ref{sec:dynamics}) 
we transfer all the atoms into the rotating $\downstate$ state. If we then ramp down $V_r$ and transform the ring trap into a simply-connected sheet trap, the phase-imprinted
$q=\ell$ vortex decays into singly-charged vortices [see Fig. \ref{fig:4}(a)]. Note however that in this case $L/N$ is no longer quantised, its exact value depending on the spatial arrangement of individual vortices \cite{Chevy:2000}. In the sheet trap the $q=3$ vortex breaks up into 3 vortices  within $1\;$s; one vortex leaves the condensate within 10 s, and the last one typically survives for about 15 s.

To quantify $L/N$ for the annular condensate, we release the atoms without letting the vortex break up in a reconnected trap. As seen in Fig.~\ref{fig:4}(b), the centrifugal barrier due to rotation of the superfluid results in a central hole in the atomic density distribution observable even after long time-of-flight (TOF) expansion \cite{Ryu:2007}.
We quantify the rotation of the cloud by fitting the radius, $R$, of the high density ring surrounding this central density hole \cite{ellipse}.

\section{Metastability and quantised decay}
\label{sec:quantisation}

\subsection{Supercurrent quantisation}

\begin{figure} [t]
\includegraphics[width=0.9\columnwidth]{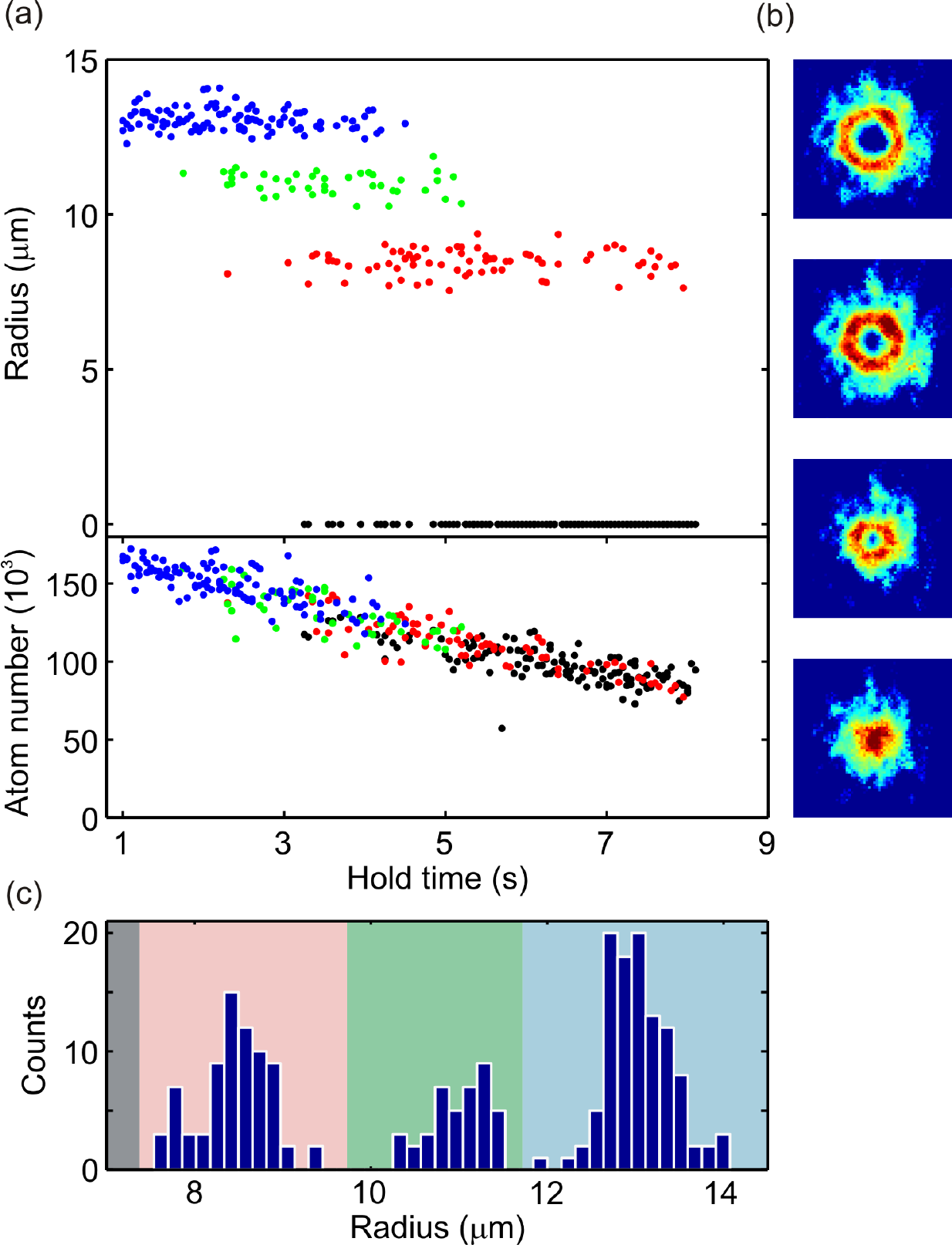}
\caption{(Color online) Quantised superflow decay. A superfluid prepared in the $q=3$ state is held in a ring trap with $V_r \approx 4\,\mu$. (a) Upper panel: radius $R$ as a function of hold time $t$. Each point is colour-coded according to the assigned $q$ value (Blue: $q=3$, Green: $q=2$, Red: $q=1$, Black: $q=0$). Lower panel: atom number $N$ versus $t$ for the same data set, with same colour code applied.   (b) TOF absorption images of the $q=3, 2, 1,$ and $0$ states. (c) High-contrast histogram of the measured $R$ values confirms that we can assign a $q$ value to each individual image with near-unity fidelity. The colour of the shaded backgrounds corresponds to our $q-$value assignments.}
\label{fig:5}
\end{figure}

The first main result of this paper is the direct experimental demonstration of the quantised nature of the supercurrent decay, shown in Fig.~\ref{fig:5} for a system initially prepared in the $q=3$ state.
In Fig.~\ref{fig:5}(a) we plot the evolution of the radius $R$ with time after the superfluid was set into rotation.
The quantisation of $R$ is strikingly obvious and we can assign a $q$ state to each individual image with $> 99\%$ fidelity.

We consider the quantisation of the supercurrent decay the primary experimental evidence for the vortex-induced phase slips as the decay mechanism.
Condensate fragmentation or collective excitations such as solitons would break the quantisation of $R$ \cite{Kanamoto:2008}, while individual particles which break away from the superflow would gradually fill up the hole in the centre of the expanded cloud; we never see any evidence of this occurring.

The broad $q=2$ and $q=1$ plateaus in Fig.~\ref{fig:5}(a) show that the intermediate $0 < q<\ell$ states are metastable even after the supercurrent decay is initiated by the first phase slip. In the analogy with a particle moving in a washboard potential [Fig.~\ref{fig:1}(a)], this corresponds to a strongly damped motion: when the system escapes from a local energy minimum it gets trapped in a new local minimum rather than rapidly decaying to $q=0$.

\subsection{Long-lived $q>1$ superflow}

The data shown in Fig.~\ref{fig:5} was obtained using a ring trap of depth $V_r \approx 4 \, \mu$.
In order to test the limits of supercurrent metastability in our setup, we also perform experiments in a very shallow trap, with $V_r$ just above the chemical potential $\mu$ \cite{ramping}.
Since the roughness of our trapping potential scales with $V_r$, reducing the trap depth to $\approx \mu$ results in the smoothest trap we can achieve. This makes the condensate density almost perfectly uniform around the ring and minimises the probability of weak links where the local $\mu$ diminishes and the phase slips are more likely \cite{Ramanathan:2011}.

\begin{figure} [bt]
\includegraphics[width=0.9\columnwidth]{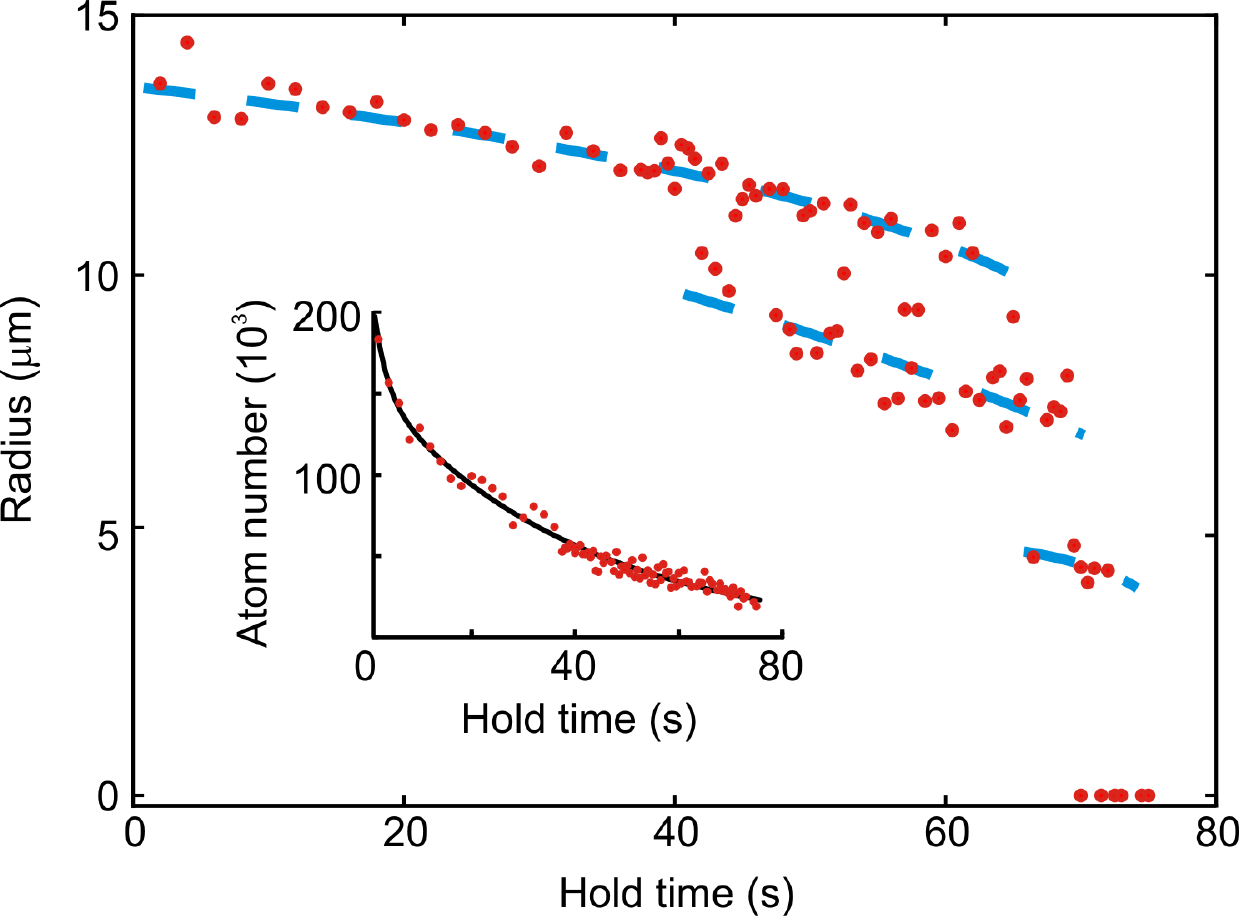}
\caption{(Color online) Long-lived $q = 3$ superflow. $R$ is plotted as a function of hold time in a shallow ring trap, showing persistent superflow for longer than a minute. The dashed lines are guides to the eye, indicating the bands of $R$ values corresponding to different $q$ states. The inset shows the decaying BEC atom number for the same data set; the solid line is a double-exponential fit to the data. }
\label{fig:6}
\end{figure}

In Fig.~\ref{fig:6} we show the evolution of $R$ for a superfluid prepared in the $q=3$ state and rotating in a shallow ring trap.
The non-zero superflow ($R > 0$) now persists for more than a minute, and decays only when the condensate itself decays significantly (see inset of Fig.~\ref{fig:6}).

The radius $R$ shows a weak dependence on the atom number $N$, making the supercurrent quantisation less striking than in Fig.~\ref{fig:5}, where the fractional variation of $N$ over the relevant timescale is much smaller. However we can still see that the $R$ values fall into distinguishable bands corresponding to $q =$ 3,  2 and 1 states. This allows us to conclude that the $q=3$ state is perfectly stable for $\sim 40\;$s and can persist for up to a minute. 
We have checked that the slow bending of the $q$ bands with time is just a consequence of the weak dependence of $R$ on the decaying $N$ (for fixed $q$), by preparing the initial $q=3$ state with deliberately reduced initial atom numbers.

In similar experiments in higher $\ell$ traps the lifetime of our BEC is shorter, but even for $\ell=10$ we still observe superflow persisting for over 20 s.

\section{Decay dynamics}
\label{sec:dynamics}

For the rest of the paper we turn to a quantitative study of the dynamics of the supercurrent decay for different superflow speeds. We first assess the critical velocity for superflow in our trap, comparing it with different theoretical calculations, and then argue that stochastic phase slips are also observed for flow below this critical velocity.

\subsection{Critical velocity}

\begin{figure} [tbp]
\includegraphics[width=0.9\columnwidth]{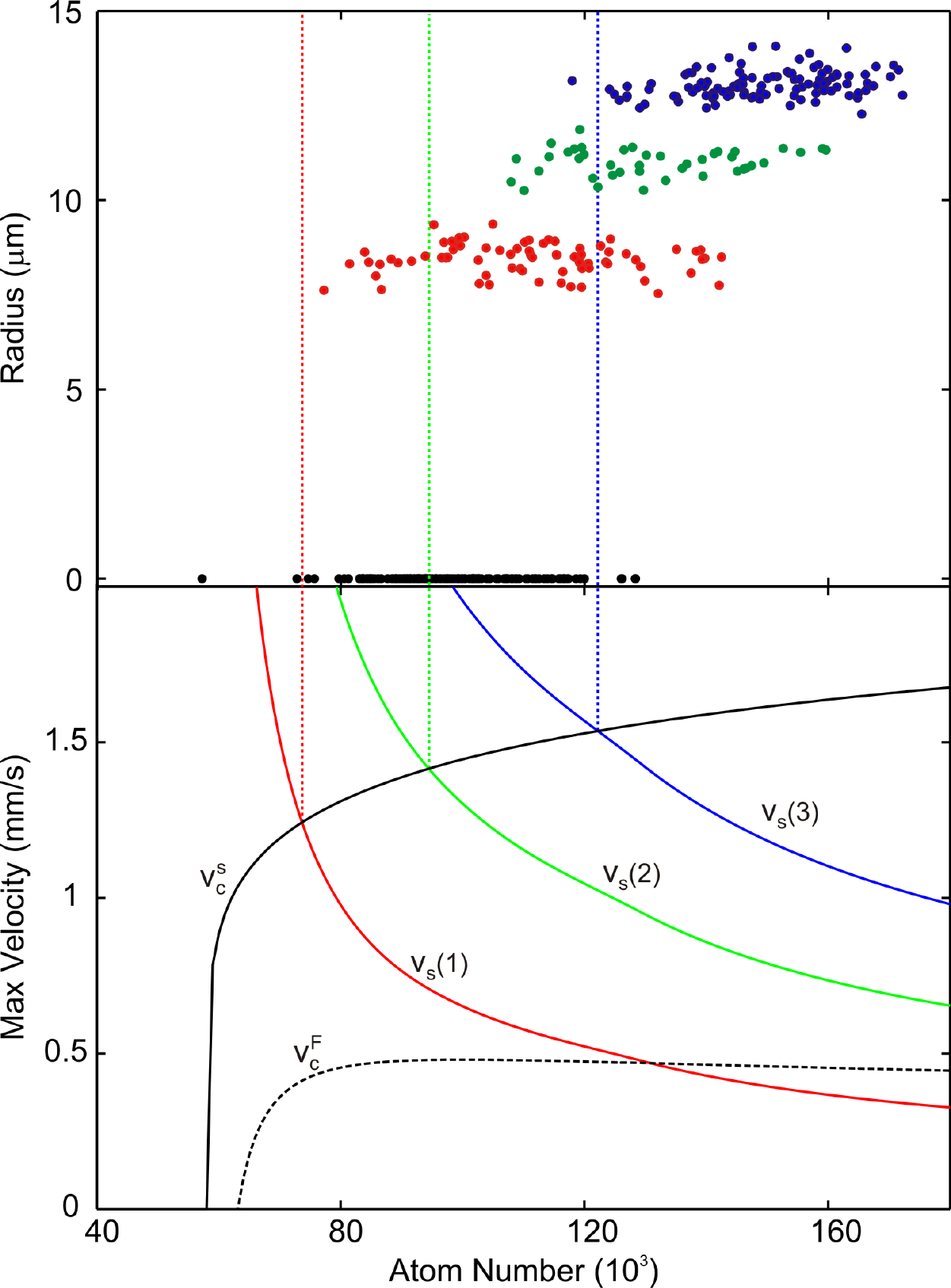}
\caption{(Color online) Comparison with numerical simulations. The top panel shows the same data as in Fig.~\ref{fig:5}, with the same colour code applied. The bottom panel shows our numerical simulations for flow speeds and critical velocities at the narrowest point in the ring (see text). The three solid coloured lines show flow speeds $v_s(q)$ for $q=3$ (blue), 2 (green) and 1 (red). The solid and dashed black curves show the calculated critical velocities $\vs$ and $\vF$, respectively. Vertical dotted lines indicate the predicted critical atom numbers $N_c^{\rm s}(q)$, defined by the intersections of  $v_s(q)$ curves and the $\vs$ curve.}
\label{fig:7}
\end{figure}

Generally, as the number of atoms in a rotating BEC slowly decays with time, superfluidity becomes less robust. Specifically, $v_s/v_c$ grows and phase slips become more likely. For comparison of our experiments with theoretical models it is convenient to eliminate the time variable and plot the observed $R$ (or equivalently $q$) values versus $N$, as shown in Fig.~\ref{fig:7}.
Here the top panel shows the same data as in Fig.~\ref{fig:5}, with the same colour code used to indicate different $q$ states. For every rotational state we see that below some $N$ the probability of observing that state sharply drops. We can thus empirically associate that ``critical" atom number $N_c(q)$ with the condition $v_s(q) = v_c$. 
The bottom panel shows the results of the numerical simulations we use to compare our measurements with the two different theoretical models outlined in Sec.~\ref{sec:theory}.

For our simulations we first use images of in-trap density distributions to assess the spatial variations in our trapping potential \cite{Ramanathan:PhD}. Next, at each point along the ring we calculate $v_s (q)$ and the two critical velocities, $\vF$ of Eq.~(\ref{eq:Feynman}) and $\vs$ of Eq.~(\ref{eq:surface}). The flow speed $v_s(q)$ is calculated under the constraints that the total circulation around the ring must be $q \hbar /m$ and the particle flux is constant along the ring. 
Finally we plot the results of our calculations for the narrowest point in the ring, where the local density and $\mu$ are lowest,  and $v_s/v_c$ is highest in both theoretical models. Note that at the point where the atom density is lowest the flow speeds up in order to conserve the particle flux. For each $q$ state the two predicted $N_c$ values, $N_c^{\rm F}(q)$ and $N_c^{\rm s}(q)$, are given by the intersections of the $v_s(q)$ curve with the two $v_c$ curves; for the $\vs$ calculations these predictions are indicated by the vertical dotted lines.

For the relevant range of $N$ values we get $\vs/\vF \sim 3$ and the $\vs$ calculation provides a much closer agreement with the data.  
For all three $q$ states the sharp drop in survival probability occurs within $\sim 15\%$ of the predicted $N_c^{\rm s}(q)$ (see vertical dotted lines).
This observation differs from that of Ref.~\cite{Ramanathan:2011}, where $v_c$ much closer to $\vF$ was observed for superflow initially prepared in the $q = \ell = 1$ state. This discrepancy warrants further investigation, but is not necessarily very surprising, given that the various differences in the trapping potentials in the two experiments are not in any way accounted for by the order-of-magnitude estimate of Eq.~(\ref{eq:Feynman}).

We note that for determining the true roughness of our trapping potential it is essential to take into account the finite resolution of our imaging system, which we model by a Gaussian point-spread function of width $\sigma$. We find the above agreement with the $\vs$ calculation by assuming $\sigma = 2.8 \; \mu$m, while we independently determine our resolution to be $2.5 \pm 0.5 \; \mu$m.

We also note that strictly speaking for the applicability of Eq.~(\ref{eq:surface}) we require the condition $\alpha = 2 (\bar{\mu}/\hbar \omega)^{2/3} \gg 1$, where $\bar{\mu}$ is now the local chemical potential at the narrowest point in the ring and for total atom number $N = N_c^{\rm s}$. For the data in Fig.~\ref{fig:7}, this condition is only marginally satisfied; $\alpha$ varies between $\approx 4$ for $q=3$ and $\approx 2$ for $q=1$. Nevertheless, the agreement with the data is still very good.

To further test the prediction of Eq.~(\ref{eq:surface}), we perform another experiment in which we again exploit the fact that the roughness of our ring potential grows with $V_r$. After preparing the BEC in the $q=3$ state we now raise $V_r$  until $q=3$ is no longer persistent but always decays to the metastable $q=2$ in $\lesssim 300\;$ms, i.e. already at the initial $N \approx 200 \times 10^3$. In other words we now measure the critical $V_r$ for a fixed $N$ and $q=3$ (see also \cite{Ramanathan:2011}). We find that the critical $V_r (\approx 6\, \mu)$ agrees to within $5\,\%$ with the  $\vs$ calculations similar to those of Fig.~\ref{fig:7}; at the critical point $\alpha \approx 4$ and $\vs \approx 3.5 \, \vF$.

\subsection{Counting statistics of stochastic phase slips}

In Fig.~\ref{fig:7} we also see evidence that some stochastic phase slips occur for $v_s < v_c$. Purely experimentally, this is directly seen in the horizontal overlaps of the different $q$ plateaus. Similar overlaps are seen in the time domain in Fig.~\ref{fig:5}(a), showing that the observed $q$ is not a deterministic function of either $t$ or $N$.
Less than $25\,\%$ of the observed overlap can be attributed to technical fluctuations in our experiments, namely shot-to-shot variations in real atom number ($\sim 3\%$) and atom-number detection ($\sim 6\%$) \cite{fluctuations}.

\begin{figure} [t]
\includegraphics[width=0.9\columnwidth]{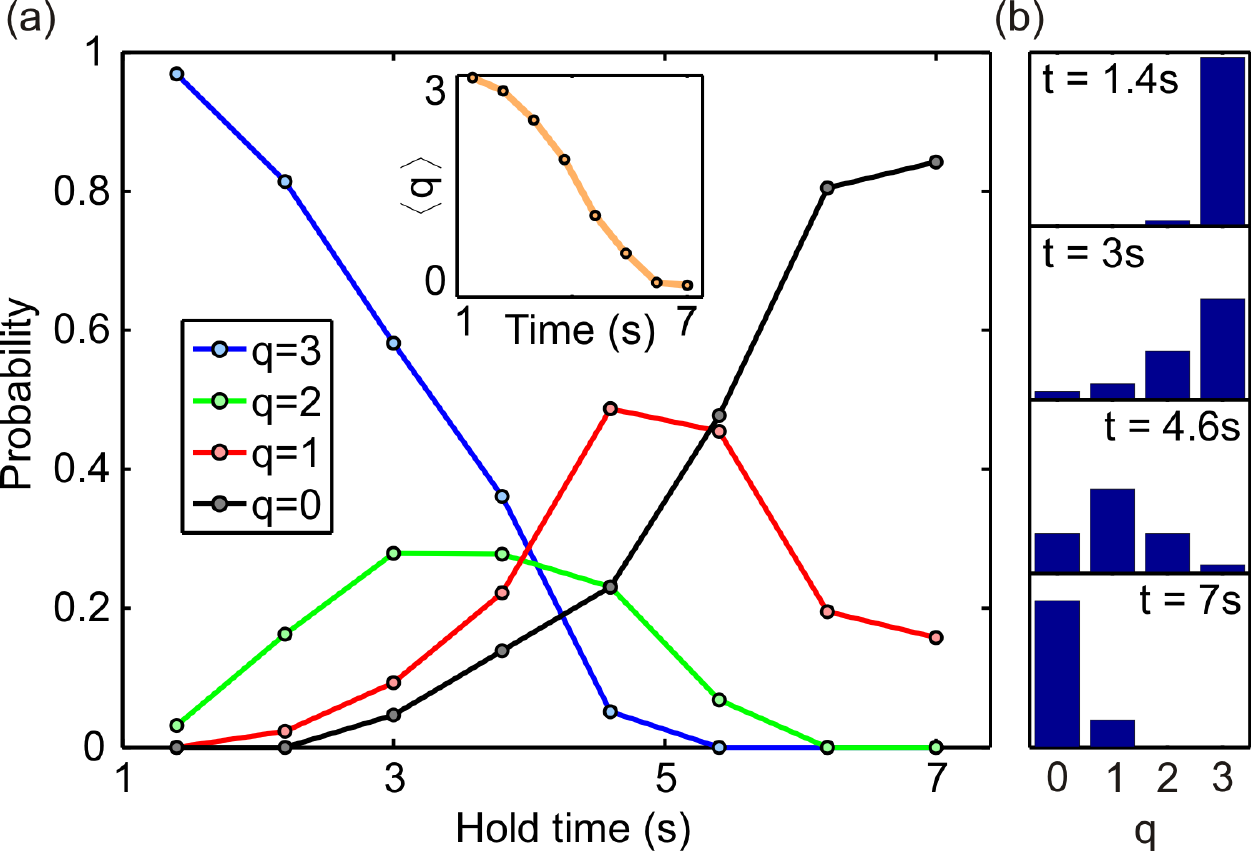}
\caption{(Color online) Counting statistics of phase slips. (a) For the data shown in Fig.~\ref{fig:5} we plot the distribution of the observed $q$ values as a function of rotation time $t$. Each data point is an average over a $0.8$ s time bin. The inset shows the smooth evolution of $\langle q \rangle$ with $t$. (b) Histograms of $q$ values for four representative rotation times.}
\label{fig:8}
\end{figure}

We therefore conclude that there exists a significant parameter space where the superflow is sub-critical but stochastic phase slips still occur on a timescale of seconds. 
In Fig.~\ref{fig:8} we show the evolution of the $q$ distribution in time, for the same data set as in Fig.~\ref{fig:5}. This in essence provides full time-resolved counting statistics of phase slips, and should be an excellent input for further theoretical modelling and understanding of the decay dynamics.
Note that this accelerating decay process is not Markovian since the phase-slip probability grows as $v_s/v_c$ increases through the gradual $N$ decay.
Also note that $\langle q \rangle$ decays smoothly with time (see inset of Fig.~\ref{fig:8}), so our demonstrated ability to experimentally resolve different $q$ states with high fidelity will be essential for further studies of phase-slip dynamics.

\section{Conclusions and outlook}
\label{sec:conclusion}

In conclusion, we have demonstrated and studied long-lived multiply-charged superflow in an annular atomic BEC. We resolve with high fidelity quantised steps in the decay of the supercurrent, which correspond to vortex-induced $2 \pi$ phase slips. 
The supercurrent decays rapidly if the flow speed reaches a critical velocity that is in agreement with numerical simulations. 
However stochastic phase slips also occur, at a much lower rate, for lower flow speeds. An important question for future work is whether these rare phase slip events occur via quantum or thermal fluctuations. Our optical setup is also suitable for spectroscopy of the excitation spectrum of an annular BEC \cite{Modugno:2006} and for studies of supercurrents in spinor condensates \cite{Smyrnakis:2009}.
Moreover, our Raman method for preparing large-$q$ rotational states can be extended to create an azimuthal gauge field \cite{Lin:2009b, Cooper:2010} and study superfluidity in continuously driven multi-component condensates.
It should also be possible to reach the regime of a narrow quasi-one-dimensional annulus, where the supercurrent decay could be fundamentally different.

\acknowledgments

We thank N. Cooper, S. Baur, M. Zwierlein, J. Dalibard, G. Campbell, E. Demler, A. Polkovnikov, D. Stamper-Kurn, E. Altman, A. Gaunt and M. Padgett for useful discussions, R. Campbell and R. Bowman for experimental assistance, and R. Fletcher for comments on the manuscript.
This work was supported by EPSRC (Grants No. EP/G026823/1 and No. EP/I010580/1) and a grant from ARO with funding from the DARPA OLE program.

%\bibliography{Ring,Quench}

\begin{thebibliography}{10}
\bibitem{LG}
Experimentally, the LG$^\ell$ mode is created by imprinting a phase winding $\ell \times 2\pi$ on a laser beam with a Gaussian modulus profile. This results in a laser mode which is a superposition of various LG$_p^\ell$ modes with fixed $\ell$ and different radial quantum numbers $p \geq 0$. For this mode the radius scales approximately linearly with $\ell$, while for the more familiar pure LG$_0^\ell$ mode the radius scales as $\sqrt{\ell}$.
\bibitem{ellipse}
Allowing for small imperfections in the ring shape, we fit an ellipse to the image and define $R$ as the geometric mean of the radii along the two principal axes.
\bibitem{ramping}
We always use the same $V_r$ for the Raman spin-up of the superfluid, and then ramp it to the desired value over 3 s. The hold time $t$ in Figs.~\ref{fig:5} and~\ref{fig:6} is measured from the end of this ramp.
\bibitem{fluctuations}
Fluctuations in $V_r$ or LG beam profile can also affect $v_s/v_c$, but we believe that these effects are negligible: the measured fluctuations of the LG beam power are  $< 2\%$, the in-trap images of the atoms suggest no shot-to-shot fluctuations in the ring profile, and we exclude slow-drift effects by comparing data taken at different times during the day.
\end{thebibliography}
%\bibliographystyle{prsty}

\end{document}